# Cloudifying the 3GPP IP Multimedia Subsystem for 4G and Beyond: A Survey [1]

Mohammad Abu-Lebdeh[1], Jagruti Sahoo[1], Roch Glitho[1], Constant Wette Tchouati[2]

[1]CIISE, Concordia University, Montreal, QC, Canada.
[2]Ericsson, Montreal, QC, Canada.

*Abstract*—4G systems have been continuously evolving to cope with the emerging challenges of human-centric and machine-to-machine (M2M) applications. Research has also now started on 5G systems. Scenarios have been proposed and initial requirements derived. 4G and beyond systems are expected to easily deliver a wide range of human-centric and M2M applications and services in a scalable, elastic, and cost efficient manner. The 3GPP IP multimedia subsystem (IMS) was standardized as the service delivery platform for 3G networks. Unfortunately, it does not meet several requirements for provisioning applications and services in 4G and beyond systems. However, cloudifying it will certainly pave the way for its use as a service delivery platform for 4G and beyond. This article presents a critical overview of the architectures proposed so far for cloudifying the IMS. There are two classes of approaches; the first focuses on the whole IMS system, and the second deals with specific IMS entities. Research directions are also discussed. IMS granularity and a PaaS for the development and management of IMS functional entities are the two key directions we currently foresee.

I. INTRODUCTION

Mobile systems have been undergoing a rather fast evolution in the recent times. 4G systems have provided increasingly higher bandwidth, lower latency, and more features to meet the more stringent requirements of human-centric and machine-to-machine (M2M) communications since their inception during the second half of the last decade. This constant innovation has paved the way for the growth of future human-centric and M2M applications and is now leading us to the 5G era.

METIS is a European project that aims to lay the foundation of the 5G concept to fulfill the requirements of the beyond-2020 connected information society and to support new usage scenarios. It identifies five service and application scenarios that 5G will have to support, namely: *amazingly fast*, *great service in a crowd*, *best experience follows you*, *super real time reliable communications*, and *ubiquitous things communicating* [1]. Several requirements are derived from these scenarios, such as much higher bandwidth, much lower latency, and much more stringent reliability and scalability than what is offered today by the evolved 4G systems. For instance, 5G systems are expected to attain 10 to 100 times higher user data rate, and 5 times lower end-to-end latency [1]. Another example is the requirement of cost efficiency, which was not a primary concern in 4G. This is certainly due to the recent emergence of new technologies such as cloud computing that can easily enable cost efficiency.

The 3GPP IP multimedia subsystem (IMS) [2] is a strong candidate for application and service provisioning in 4G and beyond because it will enable a smooth migration. It was specified as the application and service delivery platform for 3G networks and was then used at the inception of 4G as the de facto service platform. However, it does not meet all of the requirements of 4G and beyond.

Cloud computing has emerged as a paradigm for delivering computing resources (e.g., servers and storage) as a utility. It promises many benefits including elasticity, efficiency in resource usage, easy application and service provisioning, and cost reduction. It has established the foundations for the emergence of network function virtualization (NFV), which aims to transform network architectures through the implementation of network functions (e.g., IMS) in software that can run on industry standard hardware. Cloud and NFV technologies can certainly aid in tackling the IMS shortcomings when it comes to the requirements of 4G and beyond mobile and wireless systems.

There are several approaches for integrating IMS and cloud technologies. Gouveia et al. [3] illustrate these approaches by presenting scenarios in a 4G network setting. In the first group of scenarios, IMS is re-engineered using cloud technologies. In the second group, IMS is used to access applications and services implemented in clouds. In this article, "cloudifying IMS" means re-engineering IMS using cloud technologies. This corresponds to the first group of scenarios. Readers interested in the use of IMS to access applications and services implemented in the cloud can consult [4].

This article is a survey on IMS cloudification for 4G and beyond. It provides a critical review of the architectures for cloudyfying IMS that have been proposed in the literature and further discusses research directions. NFV based architectures are included in our review. The architectures proposed for IMS cloudification thus far focus on either the entire IMS system or on specific entities. We start by introducing IMS, cloud computing and NFV, also outlining the requirements of IMS cloudification for 4G and beyond. The third section reviews the architectures that focus on the entire IMS system. In the fourth section, we discuss the architecture that focuses on specific IMS entities. The fifth section focuses on research directions, and we conclude in the final section.

---

[1] This article is an extended version of a paper presented at NTMS 2014 under the title "Cloudifying the 3GPP IP Multimedia Sub-system: Why and How?"

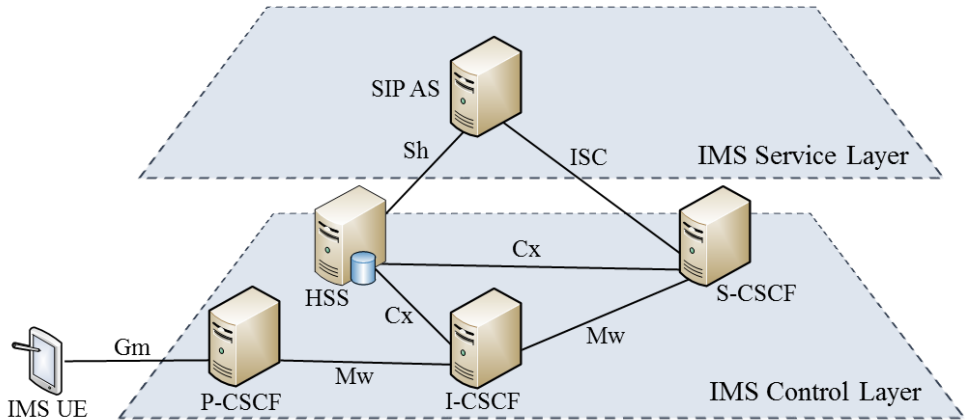

**Figure 1.** Simplified IMS architecture.

## II. BACKGROUND INFORMATION ON IMS, CLOUD COMPUTING / NFV AND REQUIREMENTS ON IMS CLOUDIFICATION

### A. IMS

IMS is an overlay control layer on top of an IP transport layer required for the seamless and robust provisioning of IP multimedia services to end-users. It is made up of a service layer and a control layer. The service layer includes application servers, such as a presence server. The key functional entity of the control layer is the call state control function (CSCF). It uses the session initiation protocol (SIP) to control multimedia functions.

Fig. 1 depicts a simplified architecture for IMS network. There are three types of CSCF: proxy-CSCF (P-CSCF), interrogating-CSCF (I-CSCF) and serving-CSCF (S-CSCF). P-CSCF is the first point of contact for the IMS user equipment (UE) within an IMS network. It acts as a stateful SIP proxy when routing SIP signaling messages going to and from an IMS UE. It is allocated to the IMS UE and does not change for the duration of the registration. I-CSCF is the first contact point for external IMS networks. It is a stateless SIP proxy that selects an S-CSCF for IMS UE and routes incoming SIP signaling messages to the selected S-CSCF. Serving-CSCF (S-CSCF) is the central node of the signaling plane of an IMS network. It acts as a stateful SIP registrar and proxy in an IMS network. As a SIP registrar, it registers IMS users and maintains the binding between the public user identity and the user profile. It also interacts with the home subscriber server (HSS) via the Cx reference point to obtain users' profiles. As a SIP proxy, S-CSCF forwards specific types of SIP messages to the appropriate application server.

HSS is another key component of the architecture. It is the central database of the mobile network that contains user-related information, such as subscription, location, and identification information. It supports the network entities' functions (e.g., mobility) and service provisioning. Several IMS functional entities at both IMS service and control layers interact with it using the diameter protocol.

The SIP application server (SIP AS) is a SIP-based server that implements the logic of IMS services. The SIP AS interacts with HSS to obtain users' profiles via the Sh reference point. An example of an IMS service is the presence service, which accepts, stores and distributes presence information via SIP messages.

The 3GPP IMS specification provides scalability through the distribution of components such as the CSCF and the HSS. However, despite this provision, scalability remains a key issue in IMS, as articulated in [5]. This is due to the fact that SIP is a text-based protocol. Signaling delay may not be sustainable when several CSCFs and application servers are deployed. In addition to the scalability issue, there is actually no provision in IMS for meeting the cost efficiency requirement of 4G and beyond mobile and wireless communications.

### B. Cloud Computing And NFV

Cloud computing has emerged as a viable delivery model for IT resources. It leverages visualization technology to enable on-demand network access to a shared pool of configurable resources (e.g., networks, servers, storage, applications, and services) with self-service provisioning and administration. It has three main service models: infrastructure as a service (IaaS), platform as a service (PaaS), and software as a service (SaaS).

IaaS offers end-users computing resources such as processing, storage, and network as a service over a network. End-users can dynamically provision and de-provision resources according to their need. Service providers use PaaS to provision applications and services that are offered as SaaS on a pay-per-use basis to end-users or other applications. PaaS eases the provisioning process by adding levels of abstraction to the infrastructure offered as IaaS. PaaS solutions vary widely in the capabilities they offer. However, they all have the basic capability to deploy applications on IaaS.

The NFV technology offers a new way to design, deploy and manage network services. It decouples network functions that are implemented in software from the underlying proprietary hardware and runs the software as applications (i.e., virtual network functions [VNFs]) on commercial off-the-shelf (COTS) hardware [6]. The shift towards software-based network functions leads to flexibility as the VNFs can be easily deployed in various locations, updated, and scaled without the need to change the hardware.

NFV was developed to benefit the networks from virtualization technology to consolidate and run VNFs on COTS hardware such as servers and switches. It promises many benefits to the Telco industry such as flexibility, openness, network services agility, and reduced capital expenditures (CAPEX) and operational expenditures (OPEX) [6].

Although related, cloud computing and NFV are different concepts. Cloud computing refers to the concept of delivering the computing resource as a service whereas NFV focuses on migrating the network functions to run on COTS hardware. However, by leveraging cloud computing, NFV can take advantage of the benefits of cloud computing and bring it to the Telco industry. The benefits include elasticity, resource efficiency, and even more reduced CAPEX and OPEX than NFV on its own.

The NFV architectural framework [8], as being standardized by the European Telecommunications Standards Institute (ETSI), is depicted in Fig. 2. It comprises NFV infrastructure (NFVI), VNFs, and NFV management and orchestration layers. NFVI provides the environment in which VNFs can execute. It provides the compute capabilities comparable to an IaaS, although usually with much more stringent performance requirements. It also supports the dynamic network connectivity between VNFs, which can be achieved by leveraging emerging technologies such as software-defined networking (SDN). The virtualized infrastructure manager performs resource management and allocation. The VNF manager handles VNF life cycle management (e.g., instantiation, scaling, and termination). The VNF orchestrator is mainly responsible for the life cycle management of the network services, which usually includes several VNF instances.

*C. Requirements*

The IMS was designed for 3G with human-centric applications in mind; however, 4G and beyond aim at catering for both human-centric applications and M2M applications (e.g., smart grid). This calls for a redesign of the IMS, and cloud computing is the ideal basis since it enables features such as scalability and efficiency in resource usage. We consider the

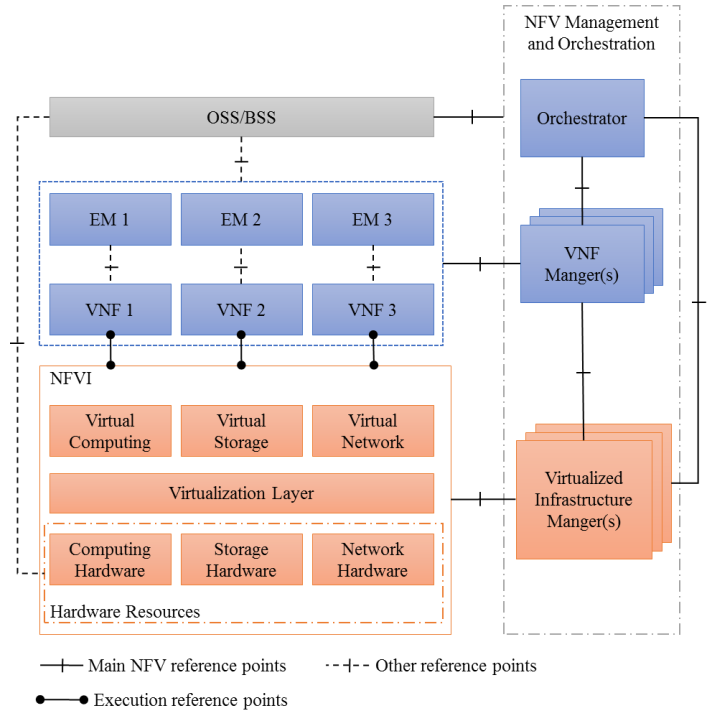

**Figure 2.** NFV architectural framework [8].

following requirements to be the most pertinent for cloudifying the IMS for 4G and beyond:

*1)* Elastic scalability: IMS today relies solely on using pre-allocated and over-provisioned functional entities to meet the expected demand peak. New capacity requires significant efforts to manually add new equipment to the system. On the other hand, a cloudified IMS should take advantage of the elasticity of the cloud to adapt dynamically to the growing or shrinking of the load requirements by adjusting the allocated resources in a fine-grained manner. Additionally, it should be able to handle smoothly a massive number of IMS UEs. Indeed, 10-100 times more devices are expected to be connected to 5G compared to today.

*2)* Latency: 4G and beyond will support a wide variety of human-centric and M2M applications that will tolerate different values of latency. Some of these applications can tolerate latencies on the order of a few seconds while others have stricter latency requirements than what exists today. For instance, teleprotection is a mission-critical application for power utilities. It includes real-time monitoring and alerting functionalities that require transferring the messages with about 8 milliseconds delay on the application layer [1]. The cloudified IMS should be able to support the applications that require different levels of latency. This includes the applications that have very strict latency requirements compared to today. It also should be able to maintain the required latency under a high load.

*3)* <u>Resource efficiency</u>: Today, IMS is installed with over-provisioning of resources to accommodate the peak demand. However, the shift towards on-demand capacity makes resource efficiency more critical, since inefficiency would be translated directly into higher running cost (i.e., OPEX) with the pay-per-use pricing model.

*4)* <u>Follow-me</u>: The basic idea behind the "follow-me" concept is that cloud services follow the end-users during their movement [7]. Mobile operators will use multiple IaaSs that are geographically distributed and interconnected [7]. IMS and IMS services could be deployed in different locations to offer better user experience. Therefore, as soon as the end-user moves, the optimal application server for providing the IMS service may change. In the future, the service should follow the end-user and should always be accessed from the application server and through the IMS functional entities which ensure the best user experience. Nowadays, P-CSCF and S-CSCF entities are allocated to the IMS UE and do not change for the duration of the registration. Through this period, end-users access their IMS services through these assigned entities. Therefore, to have service mobility in this model, the IMS UE should de-register from the assigned IMS entities and then register again which will cause service interruption.

Requirements are unfortunately often in conflict, and our proposed requirements are no exception to that tendency. Appropriate trade-offs will need to be made when new architectures are designed. Let us illustrate this by demonstrating the conflicts between elastic scalability, latency, and resource efficiency. It is clear that today's granularity level (i.e., 3GPP functional entities) is an impediment to elastic scalability. However, refining that level of granularity through the splitting of the functional entities will usually lead to an additional cost (e.g. management complexity, inter sub-functional entities communications). These costs may (or may not) offset the gains expected from the refining. In addition, the splitting may prevent latency requirements from being met. Optimal splitting, therefore, becomes the key. We further elaborate on this in the research directions section.

### III. APPROACHES THAT DEAL WITH THE ENTIRE IMS

This section reviews the approaches that focus on the whole IMS system in the light of the requirements set forth in section II.C. In these approaches, a common pool of resources is dynamically allocated to IMS functional entities. Fig. 3 provides an illustration. The physical computation, storage, and networking resources are virtualized. This allows for an IMS with a set of interacting virtual functional entities (i.e., functional entities that rely on virtualized resources). Table 1 summarizes the review findings.

*A. Virtualized IMS*

In [9], Lu et al. propose a cloud platform for the IMS core network that runs IMS entities on cloud-based virtual machines (VMs). The proposed platform supports dynamic resource allocation and disaster protection. The proposed resource allocation algorithm can dynamically allocate and de-allocate virtual central processing unit (vCPU) and memory resources to VMs according to the current workload. The algorithm aims to allow the platform to satisfy the carrier-grade response time requirement, achieve high resource utilization and reduce cost.

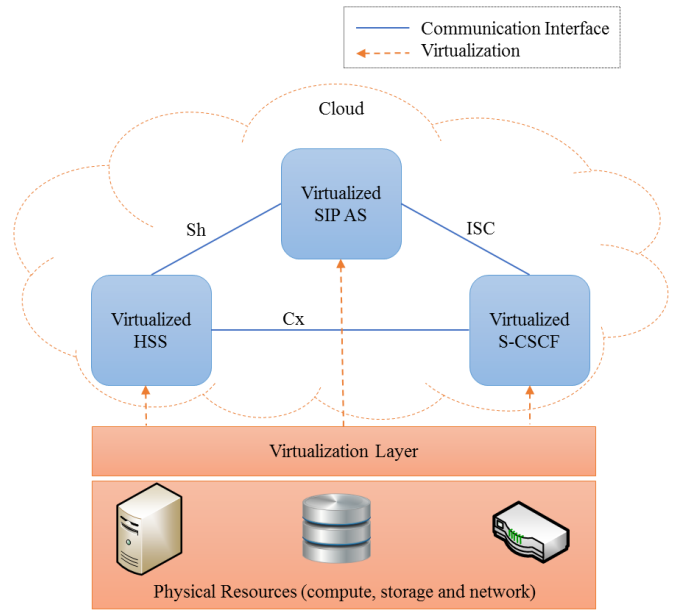

**Figure 3.** Simplified virtualized IMS.

Additionally, the algorithm assumes that each VM boots with an initially allocated vCPU and memory. Each VM also has a predefined maximum amount of vCPU and memory that can be allocated. When the resource utilization exceeds a predefined threshold, the system adds one vCPU or more memory if the VM has not reached the maximum allowed resources. If the physical machine (PM), which hosts the VM, does not have enough resources to scale the resources of the VM, the algorithm performs live migration of the VM to another PM with enough resources. The algorithm can also elastically scale the number of the active PMs in the cloud infrastructure automatically, according to the workload. It aims to achieve high resource utilization and reduce power consumption costs.

The proposed resource allocation algorithm can elastically scale IMS vertically to adapt to the workload whereas horizontal scalability is not tackled. However, the stateful architecture for many of the IMS functional entities (e.g., S-CSCF) hinders the implementation of horizontal scalability. For instance, it would be difficult to terminate an S-CSCF instance when it handles an ongoing call because this would require transferring the stored state to another S-CSCF instance. The authors propose a resource allocation algorithm to achieve high resource utilization. However, resource efficiency may not be maximal since the optimal splitting is not considered and the default splitting (i.e., IMS functional entities as defined today) is used. The authors also do not evaluate the latency achieved by their architecture. Furthermore, they do not tackle follow-me requirement. However, it remains an issue in the proposed design due to the static assignment of IMS functional entities for a specific IMS UE at the registration process.

*B. IMS as a Service*

Carella et al. [10] propose three architectures for cloud-based virtualized IMS using NFV: Virtualized-IMS, Split-IMS, and Merge-IMS. In the Virtualized-IMS architecture, each IMS functional entity is implemented as software that runs on a single VM. The interfaces with external components are not changed. The Split-IMS moves the state of the subscribers, which is maintained in many IMS entities (e.g., P-CSCF and S-CSCF), to an external functional entity called Shared-Memory. This makes the IMS entities stateless. A load balancer is positioned as an entry point for the new stateless entities to distribute the load.

Additionally, the Merge-IMS architecture groups the main four entities of IMS (i.e., P-CSCF, S- CSCF, I-CSCF, and HSS) and deploys them into one VM called IMS-VM. It introduces the IMS-Locator entity, which assigns the subscribers to a particular IMS-VM instance during the registration process. All HSS entities in IMS-VM instances share the same database to store subscriber information.

The Virtualized-IMS architecture can scale using the procedures already standardized by 3GPP to some extent. However, the scalability is limited due to the stateful architecture. The Split-IMS architecture separates functional entities' logic and state so the logic can scale easily by instantiating new stateless entities and adding them to the load balancer. However, the scalability is not in a fine-grained manner. For instance, to scale HSS, a full-fledged HSS (e.g., storage and all reference points) should be instantiated. It is also important to verify the optimality of the proposed splitting, and how it affects the performance and resource efficiency. Moreover, the Merge-IMS architecture scales by creating a new IMS-VM that has all items (i.e., full-fledged IMS). Thus, the elastic scalability is limited since it is difficult to scale in due to the granularity level (i.e., IMS-VM) and the stateful architecture.

The authors do not provide performance metrics to evaluate the architectures' latency. They also do not tackle the optimal splitting of the IMS functional entities. Therefore, it is hard to assess whether resource efficiency could be met. Although follow-me requirement has not been tackled, none of the proposed architectures can satisfy it without re-architecting the IMS.

IV. APPROACHES THAT DEAL WITH SPECIFIC IMS ENTITIES

This section reviews the approaches that focus on specific IMS entities. The main IMS entities that have attracted the attention of researchers are the HSS and the presence service. This is probably due to the fact that they are much less complex than other nodes such as the CSCF. Table 1 summarizes the review findings.

*A. HSS*

Few works propose virtualized and cloud-based HSS architectures. Yang et al. [11] propose the distribution of HSS into a resource layer and a management layer. The resource layer is implemented in the cloud and simulations are performed to demonstrate performance gains. Although their proposed solution enables an independent scaling of resource and management layers, elastic scalability is not tackled. It is also not possible to evaluate the resource efficiency as the optimal splitting is not tackled. The performed experiment shows that the latency is high. Furthermore, follow-me is not tackled in their work.

In [12], Paivarinta et al. use home location register (HLR) to evaluate whether cloud technologies can meet the carrier-grade requirements. HLR is the primary subscriber database for mobile networks up to the 3GPP release 4 standards, and today is considered a subset of HSS. The proposed architecture uses HBase NOSQL database as HLR storage and deploys it on Amazon EC2 IaaS. It utilizes the telecommunication application transaction processing (TATP) benchmarking tool to measure the performance of the HBase database under load, which is typical in telecommunications. Unfortunately, the authors do not tackle elastic scalability. They also do only discuss the storage of HLR and do not discuss the HLR application logic. Thus, it is not possible to evaluate the resource efficiency. In addition, the performed experiment shows that the latency increases proportionally to the throughput. Follow-me requirement has not been tackled.

*B. Presence service*

In [13], Belqasmi et al. propose an early architecture for a virtualized presence service for the future Internet. Although the scalability is not tackled, it is ensured through the use of presence service substrates. However, the authors do not tackle the level of granularity of the substrates. It is therefore rather difficult to assess whether the architecture could scale in a fine-grained manner and whether resource efficiency could be ensured. The latency and follow-me requirements are not tackled.

Quan et al. [14] also focus on presence service. They propose a cloud-based implementation of presence service. The Eucalyptus cloud open source software is used, and the whole presence server is deployed on a VM. The authors do not tackle elasticity scalability, optimal splitting, and follow-me. Moreover, the evaluation shows that the architecture's latency is high and increases proportionally to the throughput.

V. RESEARCH DIRECTIONS

Research on IMS cloudification has started. This section provides insightful directions for future studies. In this article, we focus on two research issues as illustrations. In the first section, we will discuss challenges related to the IMS granularity level. This discussion includes both architectural and algorithmic issues. The second section focuses on PaaS for IMS.

| **Architectures** | | **Requirements** | | | |
|---|---|---|---|---|---|
| | | *Elastic Scalability* | *Latency* | *Resource Efficiency* | *Follow-me* |
| *Virtualized IMS* | [9] | Partly | NO | Partly | NO |
| *IMS as a Service (Virtualized-IMS)* | [10] | Partly | NO | NO | NO |
| *IMS as a Service (Split-IMS)* | | Partly | NO | NO | NO |
| *IMS as a Service (Merge-IMS)* | | Partly | NO | NO | NO |
| *HSS* | [11] | NO | NO | NO | NO |
| | [12] | NO | NO | NO | NO |
| *Presence* | [13] | Partly | NO | NO | NO |
| | [14] | NO | NO | NO | NO |

**Table 1.** Summary of the evaluated approaches vs. the identified requirements for cloudifying the IMS for 4G and beyond.

*A. Reconsidering The Granularity Level of IMS*

Each IMS network functional entity as defined by 3GPP contains a set of functions as one deployable and scalable unit. These entities are often stateful which hinders elastic scalability and resiliency in the cloud. We believe it is worthy, in the cloud environment, to investigate the possibility of having finer granularity for IMS network functional entities to achieve finer control, elastic scalability, and better resiliency.

A good starting point may be to separate the functional entities' logic and data (or state). It should be noted that 3GPP has also stipulated this separation [15] primarily for data consistency purposes, but has also mentioned better scalability as a potential advantage. This brings about the challenge of leveraging cloud technologies (e.g., distributed cache) to ensure equivalent performance characteristics.

A next step will be to consider decomposing the IMS network functional entities' logic into smaller sub-functional entities, leading to finer control over the distinct functions. However, this decomposition may not be priceless. Indeed, it increases the management complexity and may have a negative impact on latency. The cloud can help to alleviate the management complexity by leveraging PaaS to automate IMS's life cycle. However, many challenges have to be addressed at the PaaS level to make it a reality. We elaborate more on this in the next section. As for the latency, placement algorithms are a potential avenue to minimize effectively the latency and cross-network traffic. For instance, the algorithms may place the related functions on VMs hosted on the same physical server, so they communicate through a virtual switch which leads to lower latency compared to the communication over the network.

The decomposition also gives rise to architectural and algorithmic challenges. At the architecture level, if the new sub-functional entities interact with each other, then there is a need to design new interfaces. The interfaces should be very lightweight to minimize the extra cost induced by the communication. On the other hand, they also need to be reliable and scalable.

At the algorithmic level, there is a need to identify the optimal granularity for the sub-functional entities that can achieve the intended benefits (if possible). A key challenge is to determine the fine-grained atomic operations of each coarse-grained IMS network functional entity (e.g., HSS and presence server), and the degree of relationship between these operations and the associated cost (e.g., memory and processing). Resource inefficiency could be translated into the cost of unused resources for given operations. This could be, for instance, translated into a graph theory problem with a weighted undirected graph formed by representing atomic operations as vertices. In this model, two vertices would be joined by an edge if they are related, and they need to communicate. It could then be solved by formulating it as an optimal clustering problem where each cluster is represented as a set of vertices. The objective would be to maximize the sum of intra-cluster communication cost, minimize the sum of resource costs of all clusters, and minimize the sum of the inter-cluster communication cost. Of course, there would be constraints such as latency. The optimal clustering problem can be solved for each coarse-grained functional entity independently. It can be shown that the optimal clustering problem is non-deterministic polynomial-time hard (NP-hard) when the number of atomic operations is large and hence requires efficient heuristics to solve it. The design of these heuristics is an important research direction. More importantly, clustering algorithms such as hierarchical clustering and K-means clustering can be modified to solve the optimal granularity problem.

*B. Towards a PaaS for IMS*

The Telco industry could leverage PaaS to deliver IMS network functional entities (e.g., CSCFs, HSS, presence, etc.) or a subset (e.g., only HSS) as SaaS services with pay-per-use pricing to end-users (i.e., IMS UE) or even to other SaaS services. The PaaS would automate the life cycle of the functional entities from deployment to management (e.g., monitoring, auto-scaling and auto-healing) and orchestration. For Telco, the PaaS would need to run on multiple, geographically distributed IaaSs that are interconnected by wide area network (WAN). This would help ensure the service continuity and reduce latency by deploying closer to end-users.

A key open issue in PaaS is the aspiration for standard language to describe the SaaS services. This language should be able to describe the structure of these services (i.e., functional entities, relationships, requirements, etc.), and their management aspects (e.g., deployment, monitoring, scaling, etc.). It should support the deployment and management across multiple IaaSs so that functional entities could be deployed at different locations. PaaS could use the services' description to automate their life cycle. Topology and orchestration specification for cloud applications (TOSCA) [16] may be a good starting point. It is standard to describe cloud applications by means of topology templates and management plans. However, the current TOSCA version (version 1.0) does not support all management aspects needed in Telco, such as monitoring.

Another research challenge is the elastic scaling of the SaaS services offered by the PaaS. These services often consist of multiple interconnected functional entities that could be distributed across multiple IaaSs. The traditional scaling approaches in PaaS usually scale the overloaded entity itself without considering the impact on other entities in the service. In Telco, these approaches would not be sufficient and efficient since there could be a need in many cases to scale and optimize other entities in the service. In fact, there is a need for new smart scaling approaches which consider the end-to-end service (i.e., all entities in the service), and is aware of service requirements (e.g., latency and resiliency) and surrounding environment (e.g., resource availability and network traffic status). These approaches should evaluate the impact of scaling and then decide accordingly what to scale, where to scale (same IaaS or across multiple IaaSs), and what to optimize aiming to meet the service requirements.

The PaaS includes management and orchestration functions (e.g., monitoring, fault management and scaling). These functions are responsible for automating the life cycle of the functional entities. However, there will be many challenges in designing them in a distributed environment. One challenge is related to the architecture and whether it is centralized or distributed. The centralized architecture is simpler, but will suffer from nontrivial latency in detecting problems and making decisions. On the other hand, the distributed architecture has lower latency. However, it is more complex and gives rise to the challenge of maintaining end-to-end service visibility. Another challenge is related to the capacity of these functions. The number of functional entities that need to be managed is changed over time as the services scale elastically. Thus, the capacity of these functions should elastically scale to adapt to system workload. This requires clear definitions of the key performance indicators need to manage the capacity.

Another open issue is network orchestration. To best of our knowledge, today IT PaaS solutions use the networks with best-effort delivery. On the other hand, quality of service (QoS) is a requirement in Telco to guarantee the performance (e.g., latency) required by the applications (e.g., multimedia applications). Indeed, Telco PaaS should leverage the IaaS network and WAN capabilities to interconnect the deployed functional entities (could be across multiple IaaSs) using transport network that meets specific requirements (e.g., latency and bandwidth). This requires that both IaaS and WAN support advanced networking capabilities (e.g., QoS) and exposes them via northbound interfaces.

## VI. CONCLUSIONS

In this article, we identified the most pertinent requirements for cloudifying the IMS for 4G and beyond. We have also reviewed the architectures proposed thus far for the cloudification of IMS using the identified requirements. These architectures are classified into two categories: the first focuses on the whole IMS system, and the second deals with specific IMS functional entity. Our evaluation has showed that the existing literature does not meet the requirements of cloudifying the IMS for 4G and beyond. Subsequently, we outlined some interesting research issues that still need to be resolved. We have discussed the possibility of decomposing IMS functional entities to achieve elastic scalability and better resiliency in cloud settings. We have also discussed the main challenges resulting from this decomposition, such as the need for new communication interfaces and optimal granularity. Furthermore, we have identified many challenges at the PaaS level. One challenge is the lack of a standard language that can describe the IMS structure and management aspects. Another challenge is the design of elastic management and orchestration functions in a distributed environment.

ACKNOWLEDGEMENT

This work is supported in part by Ericsson and the National Science and Engineering Research Council (NSERC) of Canada.

BIOGRAPHIES

Mohammad Abu-Lebdeh received his B.Sc. degree in Computer Engineering from An-Najah National University, Palestine, and M.Sc. degree in Electrical & Computer Engineering from Concordia University, Canada. He is currently pursuing his Ph.D. degree in Information & Systems Engineering at Concordia University. In the past, he worked for several years as a software engineer. His current research interests include cloud computing, service engineering, and next generation networks.

Jagruti Sahoo holds Ph.D. degree in computer science and information engineering from National Central University, Taiwan. She is currently a postdoctoral fellow at Concordia University, Canada. In the past, she has worked as postdoctoral fellow in the Department of Electrical and Computer Engineering, University of Sherbrooke, Canada. Her research interests include cloud computing, network functions virtualization, content delivery networks and wireless sensor networks.

Roch Glitho holds a Ph.D. (Tekn. Dr.) in tele-informatics (Royal Institute of Technology, Stockholm, Sweden), and M.Sc. degrees in business economics (University of Grenoble, France), pure mathematics (University of Geneva, Switzerland), and computer science (University of Geneva). He is an Associate Professor and Canada Research Chair at Concordia University in Montreal (Canada). He is also an adjunct professor at several other universities including Telecom Sud Paris (France) and the University of Western Cape (South Africa). In the past, he has worked in industry and has held several senior technical positions (e.g. senior specialist, principal engineer, expert) at Ericsson in Sweden and Canada. His industrial experience includes research, international standards setting, product management, project management, systems engineering and software/firmware design. He has also served as IEEE distinguished lecturer, Editor-In-Chief of IEEE Communications Magazine, and Editor-In-Chief of IEEE Communications Surveys & Tutorials Journal.

Constant Wette Tchouati is a Senior System Developer at Business Unit Cloud & IP, Ericsson Canada. He holds a M.Sc. in Computer Engineering from Ecole Polytechnique de Montreal (Canada), an M.Eng. In Electronics & Communication from Ecole Polytechnique de Yaounde (Cameroon) and an MBA in Management from HEC Montreal (Canada). He joined Ericsson in 2000 and has been involved in research, product development, innovation projects and new business development in different areas including Cloud architectures, Data Management and Analytics, IoT/M2M, IMS and Social Media.